\documentclass[aps,showpacs]{revtex4}
\usepackage{epsfig}
\begin{document}
\title{Multiple time scales in volatility and leverage correlations: An stochastic volatility model}
\author{Josep Perell\'o\footnote{perello@ffn.ub.es}, Jaume Masoliver\footnote{jaume@ffn.ub.es},}
\affiliation{Departament de F\'{\i}sica Fonamental, Universitat de Barcelona,\\ Diagonal, 647, 
E-08028 Barcelona, Spain}
\author{and Jean-Philippe Bouchaud\footnote{jean-philippe.bouchaud@science-finance.fr}}
\affiliation{Service the Physique de l'\'Etat Condens\'e, Centre d'\'Etudes de Saclay,\\
Orme des Merisiers, 91191 Gif-sur-Yvette Cedex, France}
\date{\today}
\affiliation{Science and Finance-CFM, 109-111 rue Victor Hugo, 92532 France}
\begin{abstract}
Financial time series exhibit two different type of non linear correlations: 
(i) volatility autocorrelations that have a very long range memory, on the order 
of years, and (ii) asymmetric return-volatility (or `leverage') 
correlations that are much shorter 
ranged. Different stochastic volatility models have been proposed in the past to 
account for both these correlations. However, in these models, the decay of 
the correlations is exponential, with a single time scale for both the volatility 
and the leverage correlations, at variance with observations. We extend the 
linear Ornstein-Uhlenbeck stochastic volatility model by assuming that the 
mean reverting level is itself random. We find that the resulting three-dimensional 
diffusion process can account for different correlation time scales. 
We show that the results are in good agreement with a century of the Dow Jones index 
daily returns (1900-2000), with the exception of crash days.
\end{abstract}
\pacs{89.65.Gh, 02.50.Ey, 05.40.Jc, 05.45.Tp}
\maketitle
\section{Introduction}
\label{intro}

The interest for financial markets models has considerably increased during the last twenty 
years. The traditional Black-Scholes Gaussian model is now well known to be inadequate; 
devising faithful and convenient models for the price processes is essential for risk control, 
derivative pricing and other trading operations. Data analysis has now become 
a mandatory step in model building, using the wealth of available tick-by-tick data or high-frequency data~\cite{dacorogna}. There is now enough data to establish, with statistical accuracy, some fundamental properties of market dynamics. Empirical observations provide a list of universal features that seem to hold in most markets, quite independently of the observation period and of the type of asset. These are the so-called stylized facts (see e.g.~\cite{cont,dacorogna2,MS}), which should be reproduced, at least to some extent, by any reasonable market model.

One of the best established stylized fact is the existence of volatility fluctuations that are long-ranged correlated in time. Empirical studies show that the volatility correlation function decays very slowly in time, and is well fitted by a power-law of the time lag with a small exponent \cite{volat1,volat2,volat3,muzy}. An exponential fit 
with a single time scale -- 
suggested by several simple stochastic volatility (SV) models 
\cite{Stein,Heston,fouquebook,schobel,perello2,Yakov} --
fails to describe the data: either the time scale is chosen to be small (a few days)
and the long time tail is completely missed, or the time scale is chosen to be
large (several hundreds of days), and the rapid initial decay is not properly accounted for. 
At least {\it two} time scales, and perhaps even three \cite{LeBaron} are needed to reproduce 
the 
volatility dynamics. Recent multifractal models \cite{Mandel,muzy,Fisher} actually postulate the existence 
of an infinity of time scales to build a self-similar volatility model with a power-law
autocorrelation function. 

Another, less well documented, effect is the negative return-volatility correlation: price 
drops are followed by an increased level of volatility. This effect was 
first noted by Black  \cite{black76} and is usually called, somewhat inappropriatly, 
the `leverage' effect, or asymmetric volatility 
correlation \cite{Beckart}. The presence of these correlations leads to skew in the 
distribution of returns
and, correspondingly, to skew in the option smiles. The effect is much stronger for 
stock indices than for individual stocks, but appears to decay on rather short time 
scales (10-20 days~\cite{bouchaud,perello2}), much shorter than the long time scale
noted above for the volatility correlation. 

Among the several SV models presented~\cite{fouquebook}, both the 
Heston~\cite{Heston,fouquebook,Yakov} and the Ornstein-Uhlenbeck~\cite{Stein,schobel,perello2} SV 
models are susceptible to contain an asymmetry parameter that generates some leverage correlations~\cite{pm}. However, as mentioned above, these models contain a {\it single} time scale which governs both the volatility
correlations and the return-volatility correlations, and cannot account for the different
temporal behaviour of the two. Multifractal models, on the other hand, have only 
been very recently extended to consider the leverage effect \cite{Pochart,muzy-new}. 
These models are however quite complex and analytical results for the unconditional
and conditional return distribution over different time scales are only partially
available \cite{muzy2,Pochart}. 

Our aim in this paper is to extend the exactly soluble Ornstein-Uhlenbeck SV model 
introduced in~\cite{Stein,schobel,perello2}, such as to reproduce
the multi-time scale dynamics reported above. In the present paper we allow for
two different time scales, but the model can be generalised to account for as many time scales
as needed, still retaining its exactly soluble nature. The basic idea is to make the 
mean reverting level of the volatility itself random and time dependent. 
The paper is divided into six sections. In section~\ref{2} we present our three-dimensional 
diffusion model. Sections~\ref{3} and~\ref{4} respectively study the leverage and volatility 
correlations. In Section~\ref{6} we apply the model to actual data, thus providing a way of 
estimating the parameters of the model. Conclusions are drawn in section~\ref{7} and some more technical details are relegated to appendices.

\section{A three-dimensional diffusion model\label{2}}

The dynamics of stock price $S(t)$ is usually modelled as a multiplicative 
diffusion process. It is commonly assumed that stock price $S(t)$ obeys the diffusion process:
\begin{equation}
\frac{dS(t)}{S(t)}=\mu dt+\sigma dW_1(t),
\label{dr}
\end{equation}
where $\mu$ is the drift, $\sigma$ is the so-called volatility and $W_1(t)$ is a 
Wiener process, {\it i.e.,} $dW_1(t)=\xi_1(t)dt$, where $\xi_1(t)$ is Gaussian white 
noise with zero-mean and unit volatility:
$$
\langle \xi_1(t) \rangle=0 \quad \mbox{and} \quad \langle \xi_1(t) \xi_1(t') 
\rangle=\delta(t-t').
$$
For the sake of simplicity we will work with the zero-mean return process defined by
\begin{equation}
dX(t)\equiv \frac{dS(t)}{S(t)}-\mu dt.
\label{zeromean}
\end{equation}
Market dynamics is then described by a simpler stochastic differential equation (SDE)
\begin{equation}
dX(t)=\sigma dW_1(t).
\label{dx}
\end{equation}

As is well-known, this model does not satisfactorily explain many properties 
of the market. A possible improvement is to assume that, 
besides the return following the diffusion process~(\ref{dr}), the volatility 
itself is a  random variable, {\it i.e.}, $\sigma=\sigma(t)$ is a stochastic process. 
This is done in the stochastic volatility models which appeared in the literature during 
late eighties~\cite{fouquebook}, following similar discrete time `ARCH' models. There are 
several ways to choose the random process describing the volatility~\cite{Heston,fouquebook}.
The simplest one consists in assuming that the volatility is also a diffusion process following 
the Ornstein-Uhlenbeck (OU) process \cite{Stein}:
\begin{equation}
d\sigma=-\alpha(\sigma-m)dt+kdW_2(t),
\label{sigsde}
\end{equation}
where $\alpha>0$, $k>0$, and $m$ are nonrandom quantities, and $dW_2(t)$ is another 
Wiener process. The quantity $1/\alpha$ is the characteristic time scale of the process and 
measures the strength of the reversion of $\sigma(t)$ to the constant $m$ which 
is the `normal' level of volatility (see below). The quantity $k$ is the amplitude 
of the volatility fluctuations and it is sometimes called `vol of vol'. As is well known, the stationary distribution of the OU process is Gaussian and this implies that the probability density does not vanish for $\sigma \to 0$. Since only $|\sigma|$ is meaningful, one has to `fold' the negative part of the Gaussian onto
the positive side with the result that the probability of $|\sigma|$ has a small hump
for $|\sigma|$ small. This and the Gaussian shape of the distribution of the 
volatility are not very realistic because the empirical distribution of the volatility is rather found to be close to a log-normal, or to an inverse Gamma distribution \cite{micchiche,bouchaudbook}. This is certainly a limitation of the present model. An interesting extension would be to consider the case where $\ln \sigma$, instead of $\sigma$, follows an OU process. Another limitation of the present model is the assumption that $dW_1(t)$ is Gaussian. Although volatility fluctuations indeed generate fat tails, the empirical frequency of very large returns (like crashes) cannot be reproduced within the class of Gaussian stochastic volatility models, and some extra `jumps' must be allowed (see the empirical results and comments in Section~\ref{6}). 

Two of us have recently studied the model represented by Eqs.~(\ref{dr})--(\ref{sigsde}) and showed~\cite{perello2} that it reproduces the leverage correlation only when the variations of volatility are anticorrelated with the variations of return. In other words, when both Wiener processes 
$W_1(t)$ and $W_2(t)$ appearing in Eqs.~(\ref{dr}) and (\ref{sigsde}) are anticorrelated:
\begin{equation}
\langle \xi_1(t) \xi_2(t') \rangle=\rho \delta(t-t'),
\label{correl0}
\end{equation}
where $\rho<0$. As mentioned above the model given by Eqs.~(\ref{dr}) 
and~(\ref{sigsde}) together with Eq.~(\ref{correl0}) reproduces very well the observed 
leverage correlation but is unfortunately unable to reproduce simultaneously 
the long range volatility autocorrelation because leverage and volatility autocorrelation 
have both approximately the same characteristic time given by $1/\alpha$. 

In the volatility process given by Eq.~(\ref{sigsde}), $m$ is a nonrandom constant 
and indicates the level to which the average volatility eventually converges as time 
increases~\cite{engle}. In order to visualize this, suppose that the process begun in the 
infinite past ($t_0\rightarrow-\infty$) which implies that at time $t$ the volatility has 
reached a stationary state, independent of initial conditions. In this situation one can 
easily see from the integration of Eq.~(\ref{sigsde}) that
\begin{equation}
\sigma(t)=m+k\int_{-\infty}^{t} e^{-\alpha(t-t')}dW_2(t').
\label{sig}
\end{equation}
From this we clearly see why $m$ can be called the normal level of volatility since 
(i) it is the stationary mean value of the volatility, $\langle\sigma(t)\rangle=m$, 
and (ii) $m$ is the value to which $\sigma(t)$ would return without any random
innovation ($k=0$).

However, there is no reason to believe that this `normal' level of volatility
is not itself slowly time evolving, with long periods where the average volatility 
is high (like stocks in the 30's), and other periods where it is smaller (like in the 60's). 
We will now relax this assumption and 
explore its consequences. We thus assume that $m=m(t)$ is a random process 
also obeying 
an OU stochastic differential equation:
\begin{equation}
dm=-\alpha_0(m-m_0)dt+k_0dW_3(t),
\label{sig0sde}
\end{equation}
where $\alpha_0>0$ and $k_0>0$, and $m_0$ is now the `true' long time average of the 
volatilty (but see the discussion in the conclusion). We therefore propose that the process describing 
the whole market dynamics is the following three-dimensional random process:
\begin{eqnarray}
&&dX(t)=\sigma(t)dW_1(t), \label{3da} \\
&&d\sigma(t)=-\alpha[\sigma(t)-m(t)]dt+kdW_2(t),\label{3db} \\
&&dm(t)=-\alpha_0[m(t)-m_0]dt+k_0dW_3(t),\label{3d}
\end{eqnarray}
where $dW_i(t)=\xi_i(t)dt$ $(i=1,2,3)$ are Wiener processes, {\it i.e.,} $\xi_i(t)$ 
are zero-mean Gaussian white noises with cross-correlation given by
\begin{equation}
\langle\xi_i(t)\xi_j(t')\rangle=\rho_{ij}\delta(t-t'),
\label{correl}
\end{equation}
where $\rho_{ij}=\rho_{ji}$ and $\rho_{ii}=1$. We also suppose that $W_3(t)$ is 
independent of $W_1(t)$ and $W_2(t)$. Hence, the correlation matrix is
\begin{equation}
(\rho_{ij})=
\left(\begin{array}{ccc}
1 & \rho & 0 \\
\rho & 1 & 0 \\
0 & 0 & 1
\end{array}\right)
\label{rho}
\end{equation}
where $-1\leq\rho\leq 1$. As is common in finance, Eqs.~(\ref{3da})-(\ref{3d}) 
are interpreted in the It\^o sense and for the rest of the paper we will 
follow the It\^o convention~\cite{perello1}. A key consequence of the It\^o 
convention is that $\sigma(t)$ and $m(t)$ are independent of any of the 
same time driving noises $dW_i(t)$ $(i=1,2,3)$. 

Note that the choice of the correlation matrix given by Eq.~(\ref{rho}) 
assumes that the variations of return and instantaneous volatility are directly correlated 
while the variation of the volatility normal level is unaffected by recent price 
changes. Although reasonable from a financial point of view, there exists a different 
limit in which the normal level of volatility is directly correlated with the return. 
This case is treated in Appendix~\ref{apC} 
where we focus on an alternative correlation matrix for which $\rho_{12}=0$ and 
$\rho_{13}=\rho$ instead of $\rho_{12}=\rho$ and $\rho_{13}=0$. We will see 
there that in this limit both the leverage and volatility correlations are
governed, on long time scales, by the `slow mode' of the model. 

In what follows we will assume that at time $t$ the entire process has reached the 
stationary state. From Eq.~(\ref{sig0sde}) we see that the expression for $m(t)$ 
in the stationary state is
\begin{equation}
m(t)=m_0+k_0\int_{-\infty}^t e^{-\alpha_0(t-t')} dW_3(t').
\label{sig0}
\end{equation}
Now the stationary expression for $m(t)$ is given by the solution of Eq.~(\ref{3db}) 
rather than by Eq.~(\ref{sig}). 
We thus have
$$
\sigma(t)=\int_{-\infty}^t[k\xi_2(t')+\alpha m(t')]e^{-\alpha(t-t')}dt,
$$
which, after using Eq.~(\ref{sig0}), reads
\begin{eqnarray}
\sigma(t)=m_0&+&k\int_{-\infty}^te^{-\alpha(t-t')} dW_2(t')
+\frac{k_0}{1-\lambda}
\int_{-\infty}^t\left[e^{-\alpha_0(t-t')}-e^{-\alpha(t-t')}\right] dW_3(t'),
\label{sigtot1}
\end{eqnarray}
where
\begin{equation}
\lambda=\frac{\alpha_0}{\alpha}.
\label{lambda}
\end{equation}
We will see later on that empirical evidence implies that $0\leq\lambda< 1$, as suggested
by the initial motivation of the model. One of the interest of this linear model is 
that one can also compute explicitely, using its Markovian nature the time evolution of the 
volatility conditioned to a given present value of the instantaneous volatility $\sigma(t_0)$ 
and of the `normal' level of the volatility $m(t_0)$. This in turn opens the way to 
obtain the conditional distribution of returns over an arbitrary horizon $T$, a quantity needed for pricing options (see \cite{Stein,schobel,perello2}).

Observe that since $m(t)$ is a Gaussian process its mean and variance read
\begin{equation}
\langle m(t)\rangle=m_0,\qquad\mbox{Var}[m(t)]=m_0^2\nu_0^2,
\label{vartheta}
\end{equation}
and the autocorrelation is
\begin{equation}
\langle m(t)m(t+\tau)\rangle=m_0^2\left(1+\nu_0^2e^{-\alpha_0\tau}\right) 
\qquad (\tau\geq 0),
\label{cortheta}
\end{equation}
where $\tau\geq 0$ and $\nu_0$ is defined by
\begin{equation}
\nu_0^2=\frac{k_0^2}{2m_0^2\alpha_0}.
\label{nu0}
\end{equation}
Likewise $\sigma(t)$ is also Gaussian and has the following averages
\begin{equation}
\langle\sigma(t)\rangle=m_0,\qquad\mbox{Var}[\sigma(t)]=
m_0^2\left(\nu^2+\nu_0^2\frac{1}{1+\lambda}\right),
\label{varsig}
\end{equation}
and autocorrelation
\begin{eqnarray}
\langle\sigma(t)\sigma(t+\tau)\rangle=m_0^2\left[1+\left(\nu^2-\frac{\lambda\nu_0^2}
{1-\lambda^2}\right)e^{-\alpha\tau}+
\frac{\nu_0^2}{1-\lambda^2}e^{-\alpha_0\tau}\right]
\label{corsigma}
\end{eqnarray}
where $\tau\geq 0$, and
\begin{equation}
\nu^2=\frac{k^2}{2m_0^2\alpha}.
\label{nu}
\end{equation}

\section{The Leverage effect\label{3}}

As recalled in the introduction, it has been known for long that volatility and returns are
negatively correlated \cite{black76}. This correlation has however not
been studied from a quantitative manner until quite recently~\cite{bouchaud}, where
the following `leverage' correlation function was studied: 
\begin{equation}
{\cal L}(\tau)\equiv
\frac{1}{Z}\langle dX(t+\tau)^2dX(t)\rangle
\label{leverage}
\end{equation}
where $dX(t)$ is the zero-mean return and
\begin{equation}
Z=\langle dX(t)^2\rangle^2
\label{Z}
\end{equation}
is a convenient normalization coefficient. A large amount of daily returns for both market 
indices and share prices were studied in~\cite{bouchaud}, with the finding that,
\begin{equation}
{\cal L}(\tau) \approx \cases{-Ae^{-\Gamma \tau}, &if $\tau>0$;\cr
0, &if $\tau<0$;}
\label{bouchaudlev}
\end{equation}
$(A,\Gamma>0)$. Hence, there is an exponentially decaying negative correlation between 
future volatility and past returns changes. No correlation is found between past 
volatility and future price changes (except perhaps for very short time lags
$|\tau|$). In this way, a clear causality of the leverage effect is established, 
whereas there are conflicting claims on this issue in the literature~\cite{Beckart}.
For individual stocks, the coefficient $A$ is found to be close to $2$, a value 
expected on the basis of simple `retarded' effect, where the price changes are locally
{\it additive}, rather than multiplicative random variables~\cite{bouchaud}. For indices,
however, $A$ is much stronger ($A \approx 20$), and the retarded interpretation fails. 
The coefficient $\Gamma$, on the other hand, defines 
the inverse relaxation time of the leverage correlation, and is found to be on the 
order of ten days for stock indices. 

Let us study the leverage correlation within the framework of the three-dimensional 
diffusion model presented in Section~\ref{2}. First, from Eq.~(\ref{3da}), we calculate 
the correlation
\begin{equation}
\langle dX(t)dX(t+\tau)^2\rangle=\langle \sigma(t)dW_1(t)\sigma(t+\tau)^2dW_1(t+\tau)^2\rangle.
\end{equation}
Following the It\^o convention, if $\tau\geq 0$ then $dW_1(t+\tau)$ is uncorrelated 
with the rest of stochastic variables. Thus, taking into account that 
$$
\langle dW_1(t+\tau)^2\rangle=dt,
$$
we have
$$
\langle dX(t)dX(t+\tau)^2\rangle=\langle \sigma(t)dW_1(t)\sigma(t+\tau)^2\rangle dt
\qquad (\tau>0).
$$
Otherwise, if $\tau\leq 0$, then $dW_1(t)$ is uncorrelated with the remaining variables. 
Hence, taking into account that $\langle dW_1(t)\rangle=0$, we get
$$
\langle dX(t)dX(t+\tau)^2\rangle=0 \qquad (\tau\leq 0).
$$
Therefore,
\begin{equation}
\langle dX(t)dX(t+\tau)^2\rangle=
\langle \sigma(t)\sigma(t+\tau)^2dW_1(t)\rangle \mbox{H}(\tau)dt,
\label{corl}
\end{equation}
where
\begin{equation}
\mbox{H}(\tau)=\cases{1, &if $\tau>0$;\cr
0, &if $\tau\leq 0$;}
\label{heaviside}
\end{equation}
is the Heaviside step function. Now the Novikov theorem allows us to write 
(see Appendix \ref{apA})
\begin{equation}
\langle dX(t)dX(t+\tau)^2\rangle=
2\rho ke^{-\alpha\tau}\mbox{H}(\tau)\langle \sigma(t)\sigma(t+\tau)\rangle dt^2,
\label{cornov}
\end{equation}
where $\langle \sigma(t)\sigma(t+\tau)\rangle$ is given by Eq.~(\ref{corsigma}).

On the other hand, as a consequence of the It\^o convention, processes $\sigma(t)$ 
and $dW_1(t)$ are independent. Therefore from Eqs.~(\ref{3da}) and~(\ref{Z}) we see that
$$
Z=\langle dX(t)^2\rangle^2=\langle \sigma(t)^2\rangle^2 \langle dW_1(t)^2\rangle^2 =
\langle \sigma(t)^2\rangle^2 dt^2.
$$
Taking into account Eq.~(\ref{varsig}), $Z$ reads
\begin{equation}
Z=m_0^4\left(1+\nu^2+\frac{\nu_0^2}{1+\lambda}\right)^2dt^2,
\label{z}
\end{equation}

Finally, the substitution of Eqs.~(\ref{corsigma}), (\ref{cornov}) and (\ref{z}) into
Eq.~(\ref{leverage}) yields
\begin{equation}
{\cal L}(\tau)=-\mbox{H}(\tau)A(\tau)e^{-\alpha\tau},
\label{lev}
\end{equation}
where
\begin{equation}
A(\tau)\equiv -\frac{2\rho\nu\sqrt{2\alpha}}{m_0\left(1+a+b\right)^2}
\left[1+a e^{-\alpha\tau}+ b e^{-\alpha_0\tau}\right],
\label{levA}
\end{equation}
with:
\begin{equation}
a=\nu^2-\frac{\lambda\hat{\nu}_0^2}{1-\lambda},\qquad
b=\frac{\hat{\nu}_0^2}{1-\lambda},
\label{ab}
\end{equation}
and
\begin{equation}
\hat{\nu}_0^2\equiv\frac{\nu_0^2}{1+\lambda}.
\label{hatnu}
\end{equation}
Note that Eq.~(\ref{lev}) is not very different from Eq.~(\ref{bouchaudlev}) above:
the dominant time decay is fixed by the short time scale of the volatility process $1/\alpha$. 
As intuitively expected, the correlation coefficient $\rho$ between $dW_1$ (returns) 
and $dW_2$ (volatility changes) needs to be negative if we want to describe the observed 
anticorrelation.

\section{Volatility correlation\label{4}}

The correlation for the volatility $\sigma(t)$ has been already given by Eq.~(\ref{corsigma}). 
Nevertheless, in order to study the different time regimes appearing in the correlation, it 
is convenient to deal with a slightly different form of correlation. We will thus work with 
the function
$$
\mbox{Corr}\left[\sigma(t),\sigma(t+\tau)\right]=
\frac{\langle\langle\sigma(t)\sigma(t+\tau)\rangle\rangle}{\sqrt{\mbox{Var}[\sigma(t)]}
\sqrt{\mbox{Var}[\sigma(t+\tau)]}}
$$
where $\langle\langle\sigma(t)\sigma(t+\tau)\rangle\rangle$ is the second cumulant of the 
volatility process defined by
$$
\langle\langle\sigma(t)\sigma(t+\tau)\rangle\rangle=\langle\sigma(t)\sigma(t+\tau)\rangle-
\langle\sigma(t)\rangle\langle\sigma(t+\tau)\rangle.
$$
From Eqs~(\ref{varsig}) and~(\ref{corsigma}) we have
\begin{equation}
\mbox{Corr}\left[\sigma(t),\sigma(t+\tau)\right]=M[ae^{-\alpha\tau}+be^{-\alpha_0\tau}],
\label{corcoef}
\end{equation}
where
\begin{equation}
M=\frac{1}{\nu^2+\hat{\nu}_0^2},
\label{M}
\end{equation}
and $\hat{\nu}_0^2$ is given by Eq.~(\ref{hatnu}).

\begin{figure}[t,b]
\epsfig{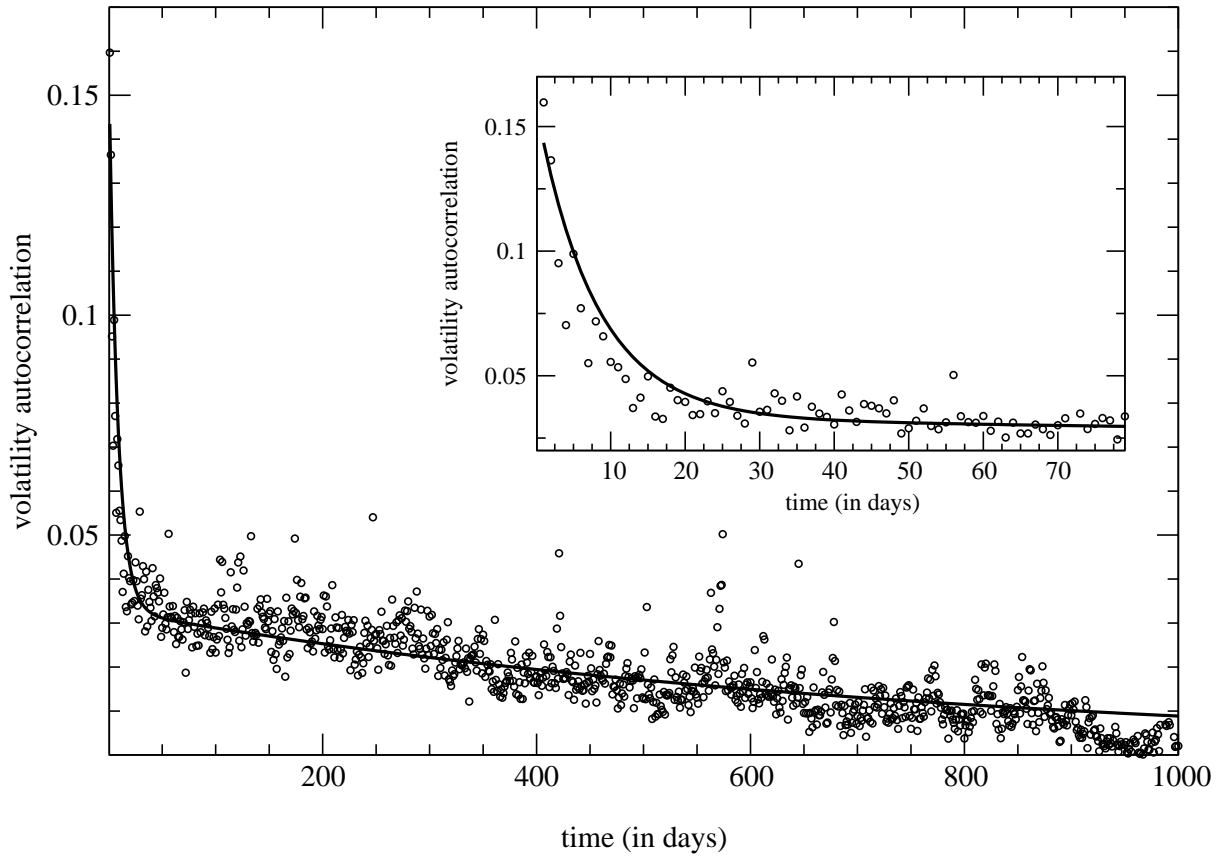}
\caption{The autocorrelation of $\Delta X(t)^2$ for the Dow-Jones daily index. We also plot 
their theoretical correlations given by Eq.~(\ref{volcor3}) with the parameters estimated 
from the fit with $\alpha=0.1 \mbox{ days}^{-1}$, $\alpha_0=1.3\times 10^{-3} 
\mbox{ days}^{-1}$, $a=0.14$ and $b=0.04$. The inset is a zoom of the correlation focussing 
on the range from 1 to 80 days.}
\label{corvol}
\end{figure}

If we assume that $\alpha_0\ll \alpha$ the volatility correlation~(\ref{corcoef}) 
has two different regimes according to $\tau$ being small or large. 
\begin{itemize}

\item Suppose that $\tau$ is small and such that the following two conditions meet
simultaneously
$$
\alpha_0\tau\ll 1\quad\mbox{and}\quad\alpha\tau\sim 1.
$$
In this case, the characteristic time scale of volatility process given by $1/\alpha$ 
(see Eq.~(\ref{3db})) will dominate and the correlation reads
\begin{equation}
\mbox{Corr}\left[\sigma(t),\sigma(t+\tau)\right]\sim M(b+ae^{-\alpha\tau}),
\qquad(\alpha_0\tau\ll 1).
\label{corcoef1}
\end{equation}

\item  Suppose now that $\tau$ is large and such that
$$
\alpha\tau\gg 1\quad\mbox{and}\quad\alpha_0\tau\sim 1.
$$
In this case the characteristic time of the normal level of volatility $m(t)$ 
(see Eq.~(\ref{3d})) is the dominant one and we see from Eq.~(\ref{corcoef}) that
\begin{equation}
\mbox{Corr}\left[\sigma(t),\sigma(t+\tau)\right]\sim Mbe^{-\alpha_0\tau},
\qquad(\alpha\tau\gg 1).
\label{corcoef2}
\end{equation}
\end{itemize}
Note that both approximations are consistent since we have assumed that $\alpha_0\ll\alpha$. 
This implies that $\lambda\ll 1$ ({\it cf.} Eq~(\ref{lambda})) is very small which, in turn,  
simplifies the form of the parameters $M$, $a$, and $b$. Indeed, now we can write
\begin{equation}
M=(\nu^2+\nu_0^2)^{-1}+\mbox{O}(\lambda),
\label{Mapprox}
\end{equation}
and
\begin{equation}
a=\nu^2+\mbox{O}(\lambda),\qquad b=\nu_0^2+\mbox{O}(\lambda).
\label{abapprox}
\end{equation}
Equations~(\ref{corcoef1}) and~(\ref{corcoef2}) show respectively the correlation behavior 
for small and large time $\tau$. The long range correlation for the volatility is governed by 
$\alpha_0$, that is, as expected, by the relaxation process of the volatility to its 
long time average $m_0$. 
This behavior is different from the leverage correlation which has a faster time decay 
$\alpha^{-1}$
that sets the `fast' reversion of the volatility. 
Notice that this is consistent with empirical observations~\cite{cont,engle,LeBaron} 
(see also Fig.~\ref{corvol} and discussion in the next section).

At this stage, we should remind that the actual evaluation of the volatility $\sigma(t)$ 
is very difficult, since the volatility itself is not observed. In practice one can measure 
a {\it noisy} measure of the instantaneous volatility from: 
\begin{equation}
|\sigma(t)| \approx  \sqrt{\left[X(t+\Delta t)-X(t)\right]^2/\Delta t},
\label{instvoldef}
\end{equation}
where $X(t)$ is the (undrifted) log-price defined above. Therefore, only $|\sigma(t)|$
can be extracted from data, and this is in principle insufficient to estimate (\ref{corcoef}). 
Another possible expression for the empirical autocorrelation of the volatility that is more
directly observable is: 
\begin{equation}
\mbox{Corr}\left[dX(t)^2,dX(t+\tau)^2\right]=\frac{1}{Y}\langle \langle dX(t)^2dX(t+\tau)^2 
\rangle \rangle,
\label{volcor}
\end{equation}
where $Y^2=\mbox{Var}[dX(t)^2]\mbox{Var}[dX(t+\tau)^2]$ and
$$
\langle \langle dX(t)^2dX(t+\tau)^2 \rangle \rangle=\langle dX(t)^2dX(t+\tau)^2\rangle-\langle
dX(t)^2\rangle\langle dX(t+\tau)^2\rangle,
$$
is the second cumulant associated with $dX(t)^2$. This correlation function is particularly 
interesting because it is directly related to the kurtosis of the terminal distribution
of price changes, a quantity which is in turn related to the curvature of volatility smiles
\cite{bouchaudbook}.

In Appendix~\ref{apB} we show that within
the present Gaussian model, the
empirical correlation coefficient given by Eq.~(\ref{volcor}) can be written in terms of the
volatility autocorrelation as
\begin{equation}
\mbox{Corr}\left[dX(t)^2,dX(t+\tau)^2\right]=\frac{\langle\sigma(t)\sigma(t+\tau)\rangle^2-
\langle\sigma(t)\rangle^4}{4\langle\sigma(t)^2\rangle^2-3\langle\sigma(t)\rangle^4},
\label{volcor2}
\end{equation}
and from Eqs.~(\ref{varsig}) and (\ref{corsigma}) we explicitly write
\begin{eqnarray}
\mbox{Corr}\left[dX(t)^2,dX(t+\tau)^2\right]=
N\biggl[a\bigl(2+ae^{-\alpha\tau}\bigr)e^{-\alpha\tau}&+&
b\bigl(2+be^{-\alpha_0\tau}\bigr)e^{-\alpha_0\tau}
+2abe^{-(\alpha+\alpha_0)\tau}\biggr]
\label{volcor3}
\end{eqnarray}
where $a$ and $b$ are given by Eq.~(\ref{ab}) and
\begin{equation}
N=\left[1+8(\nu^2+\hat{\nu}_0^2)+4(\nu^2+\hat{\nu}_0^2)^2\right]^{-1}.
\label{N}
\end{equation}
Observe that correlation~(\ref{volcor3}) consists of a linear combination of decaying 
exponentials with arguments $\alpha_0\tau$, $\alpha\tau$, $2\alpha_0\tau$, $2\alpha\tau$, 
and $(\alpha_0+\alpha)\tau$. We study some asymptotic limits for the 
correlation~(\ref{volcor3}) as done above with $\sigma(t)$. We again consider the case 
when $\alpha_0\ll\alpha$ ({\it i.e.,} $\lambda\ll 1$) and study the short and long $\tau$ 
behavior. For small $\tau$, when $\alpha_0\tau$ is small but $\alpha \tau$ is finite and 
not too small, we have
\begin{equation}
\mbox{Corr}\left[dX(t)^2,dX(t+\tau)^2\right] \approx N\left[(2+b)b+
a\left(2+2b+ae^{-\alpha\tau}\right)e^{-\alpha\tau}\right],\quad(\alpha_0\tau\ll 1).
\label{volcor4}
\end{equation}
In this regime, we only have exponentials with $\alpha$ as it happened with the sigma 
autocorrelation ({\it c.f.} Eq.~(\ref{corcoef1})). For large $\tau$, the correlation 
is again dominated by the exponential with characteristic time scale $\alpha_0$ and reads
\begin{equation}
\mbox{Corr}\left[dX(t)^2,dX(t+\tau)^2\right]\approx
Nb\left(2+be^{-\alpha_0\tau}\right)e^{-\alpha_0\tau}
\qquad (\alpha\tau\gg 1).
\label{volcor5}
\end{equation}
Observe that pairs of Eqs.~(\ref{corcoef1})--(\ref{corcoef2}) and 
Eqs.~(\ref{volcor4})--(\ref{volcor5}) contain the behavior that is empirically observed 
in financial markets. Although Eq.~(\ref{volcor3}) has a much more complicated 
form, it still contains the two desired regimes 
for large and small $\tau$ as we have discussed above.

\begin{figure}[t]
\epsfig{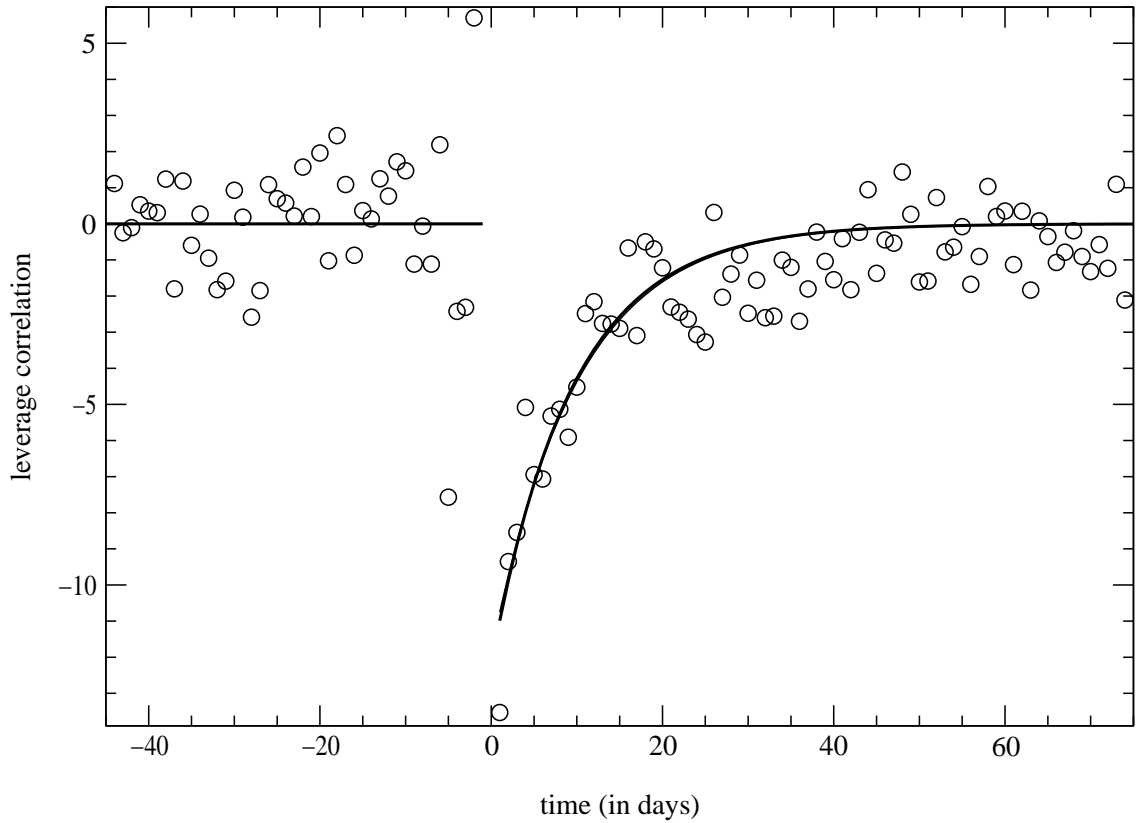}
\caption{The leverage effect in the Dow-Jones daily index. We plot the leverage 
function ${\cal L}(\tau)$ for the Dow-Jones index from 1900 until 2000. We also plot 
a fit with the leverage function~(\ref{lev}) with $\alpha=0.1 \mbox{ days}^{-1}$, 
$a=0.14$, $b=0.04$, and $m=1.19 \times 10^{-2} \mbox{ days}^{-1/2}$ finding that 
$\rho=-0.48$ (see Table~\ref{corestim}).}
\label{djleverage}
\end{figure}

\section{Estimation of parameters\label{6}}

Any acceptable market model must be easily calibrated to reproduce the dynamics of the market. 
In other words, the model should be supplemented by a systematic methodology for estimating 
its parameters. In our model we have six parameters 
to estimate. These parameters are $\rho$, $k$ and $k_0$ (or $\nu^2$ and $\nu_0^2$), $\alpha$ 
and $\alpha_0$, and $m_0$ (see Eqs.~(\ref{3da})--(\ref{3d})).

Let us discuss the way to estimate these parameters. We will proceed in a very similar 
way to the case given in Ref.~\cite{perello2} and perform the empirical estimation of 
the Dow-Jones index daily returns by approximating $dX$ by 
$\Delta X\equiv X(t+\Delta t)-X(t)$, {\it i.e.},
$$
dX(t)\approx X(t+\Delta t)-X(t),
$$
where $\Delta t=1$ day. From Eqs.~(\ref{varsig}) and~(\ref{varsig2}) and 
taking into account that $\sigma(t)$ is independent of $dW_1(t)$, and that 
$dW_1(t)$ is Gaussian, we have
\begin{eqnarray}
\mbox{Var}[\Delta X]&=&\langle \sigma(t)^2\rangle \Delta t= m_0^2(1+\nu^2+\hat{\nu}_0^2)
\Delta t,\label{varx}
\\
\mbox{Var}[\Delta X^2]&=&\left[3\langle \sigma(t)^4\rangle-\langle \sigma(t)^2\rangle\right]
\Delta t^2 =2m_0^4\left[4(1+\nu^2+\hat{\nu}_0^2)^2-3\right] \Delta t^2,\nonumber
\end{eqnarray}
where $\nu^2=k^2/(2\alpha m_0^2)$ and $\hat{\nu}_0^2=k_0^2/(2\alpha_0 m_0^2(1+\lambda))$.
From these equations we get
\begin{equation}
\frac{1}{(1+\nu^2+\hat{\nu}_0^2)^2}=\frac43-\frac16 \frac{\mbox{Var}[\Delta X^2]}
{\mbox{Var}[\Delta X]^2}.
\label{varestim}
\end{equation}
Hence, we can estimate the values of $(\nu^2+\hat{\nu}_0^2)$ once we know the
empirical values of the daily variances of $\Delta X$ and $\Delta X^2$. In
a second step, $m_0$ is estimated from the knowledge of 
$(\nu^2+\hat{\nu}_0^2)$ and the empirical value of $\mbox{Var}[\Delta X]$. 
In Table~\ref{momestim}, we briefly report these operations and give the 
corresponding estimation of $(\nu^2+\hat{\nu}_0^2)$ and $m_0$ for the Dow-Jones 
index time-series from 1900 until 2000.

\begin{table}[tbp]
\caption{Estimate from return variances. We estimate parameters of the model 
from Dow-Jones historical daily returns from 1900 to 2000. We use the 
identity~(\ref{varestim}) and take the variance~(\ref{varx}) to derive the 
estimated quantities $(\nu^2+\hat{\nu}_0^2)$ and $m_0$.}
\begin{center}
\begin{tabular}{ccc}
\hline
\hline
\\
{\footnotesize Estimators} & {\footnotesize Dow-Jones daily return} & 
{\footnotesize Theoretical values} \\
\cline{1-3}
\\
$\mbox{Var}[\Delta X(t)]$ & $1.68 \times 10^{-4}$ & $m_0^2(1+\nu^2+\hat{\nu}_0^2) \Delta t$
\\
$\mbox{Var}\left[\Delta X(t)^2\right]$ & $10.5 \times 10^{-8}$ & $2m_0^4\left
[4(\nu^2+\hat{\nu}_0^2)^2-3\right] \Delta t^2$ \\ \\
\multicolumn{2}{l}{\footnotesize Parameter estimates} 
\\
$(\nu^2+\hat{\nu}_0^2)$ & 0.18 & \\
$m_0$ & $18.9 \% \ \mbox{year}^{-1/2}$ \\ \\
\hline \hline
\end{tabular}
\end{center}
\label{momestim}
\vspace*{13pt}
\end{table}

Since $(\nu^2+\hat{\nu}_0^2)$ is always positive and $(1+\nu^2+\hat{\nu}_0^2)^2\geq 1$,
we see from Eq.~(\ref{varestim}) that
$$
2\leq\frac{\mbox{Var}[\Delta X^2]}{\mbox{Var}[\Delta X]^2}<8,
$$
and the kurtosis defined by 
$$
\kappa=\frac{\mbox{Var}[\Delta X^2]}{\mbox{Var}[\Delta X]^2}-2
$$
has the bounds
\begin{equation}
0\leq \kappa<6.
\label{bounds}
\end{equation}
As expected, volatility fluctuations induce some excess kurtosis, but this excess kurtosis
cannot take arbitrarily high values. For the Dow-Jones index data and after 
taking away `crash days' of amplitude greater than $12 \%$, we get a kurtosis of 
$\kappa \approx 1.7$ which is consistent with the requirement~(\ref{bounds}). 
The removal of these outliers only represents 11 data points, a very small fraction of 
the total amount (more than 27000 sample data points). However, as many authors have 
reported~\cite{Lux,Longin,Stanley}, the total kurtosis is found to be larger, 
perhaps even infinite (in particular for higher -- intraday -- frequencies). 
This is a limitation of the present model for describing higher frequency movements 
where one should eventually take into account a non Gaussian character for $dW_1$ (see the discussion in ~\cite{bouchaudbook} and Refs.~\cite{mmp,mmw})

\begin{table}
\caption{Parameter estimation calculated from the volatility autocorrelation~(\ref{volcor3}) and the leverage correlation~(\ref{lev}). We estimate $\alpha$, $\hat{\nu}_0^2$, and $\nu^2$ from the empirical volatility autocorrelation of Dow-Jones stock index plotted in Fig.~\ref{corvol}. We estimate the parameters $\rho$ and check $\alpha$ from the fit of the Dow-Jones stock index data leverage plotted in Fig.~\ref{djleverage}.}
\begin{center}
\begin{tabular}{lr}
\hline \hline \\
{\footnotesize Estimators} & {\footnotesize Dow-Jones data estimation} \\
\cline{1-2}
\\
\multicolumn{2}{l}{Volatility autocorrelation}
\\
$\alpha$ & $ 0.1 \mbox{ days}^{-1}$ \\
$\alpha_0$ & $1.3\times 10^{-3} \mbox{ days}^{-1}$ \\
$a\simeq\nu^2$ & 0.14 \\
$b\simeq\nu_0^2$ & 0.04 \\ \\

$\lambda=\alpha_0/\alpha$ & $1.3 \times 10^{-2}$ \\
$k=m_0\sqrt{2\alpha \nu^2}$ & $2.0 \times 10^{-3} \mbox{ days}^{-1}$ \\
$k_0=m_0\sqrt{2\alpha_0\nu_0^2}$ & $1.2\times 10^{-4} \mbox{ days}^{-1}$ \\
\\
\multicolumn{2}{l}{Leverage}
\\
${\cal L}(0^+)$ & -11.9 \\
$\rho$ & -0.48
\\
\\
\hline \hline
\end{tabular}
\end{center}
\label{corestim}
\vspace*{13pt}
\end{table}
\begin{figure}[t]
\centerline{\epsfig{file=simulation.eps,width=16cm}}
\caption{Dow-Jones index and simulation paths over 10000 days which corresponds to 
approximately 40 years. Top figure shows the Dow-Jones index daily return changes over 10000 trading days and compare it with several simulations. From top to bottom we show random path of our three-dimensional stochstic volatility model~(\ref{3da})-(\ref{3d}), the path of the two-dimensional approach~(\ref{dx})-(\ref{sigsde}) and the Wiener process~(\ref{dx}) assuming constant $\sigma$. Parameters are given in Tables~\ref{momestim}
and~\ref{corestim}.}
\label{comp}
\end{figure}
\begin{figure}[t]
\centerline{\epsfig{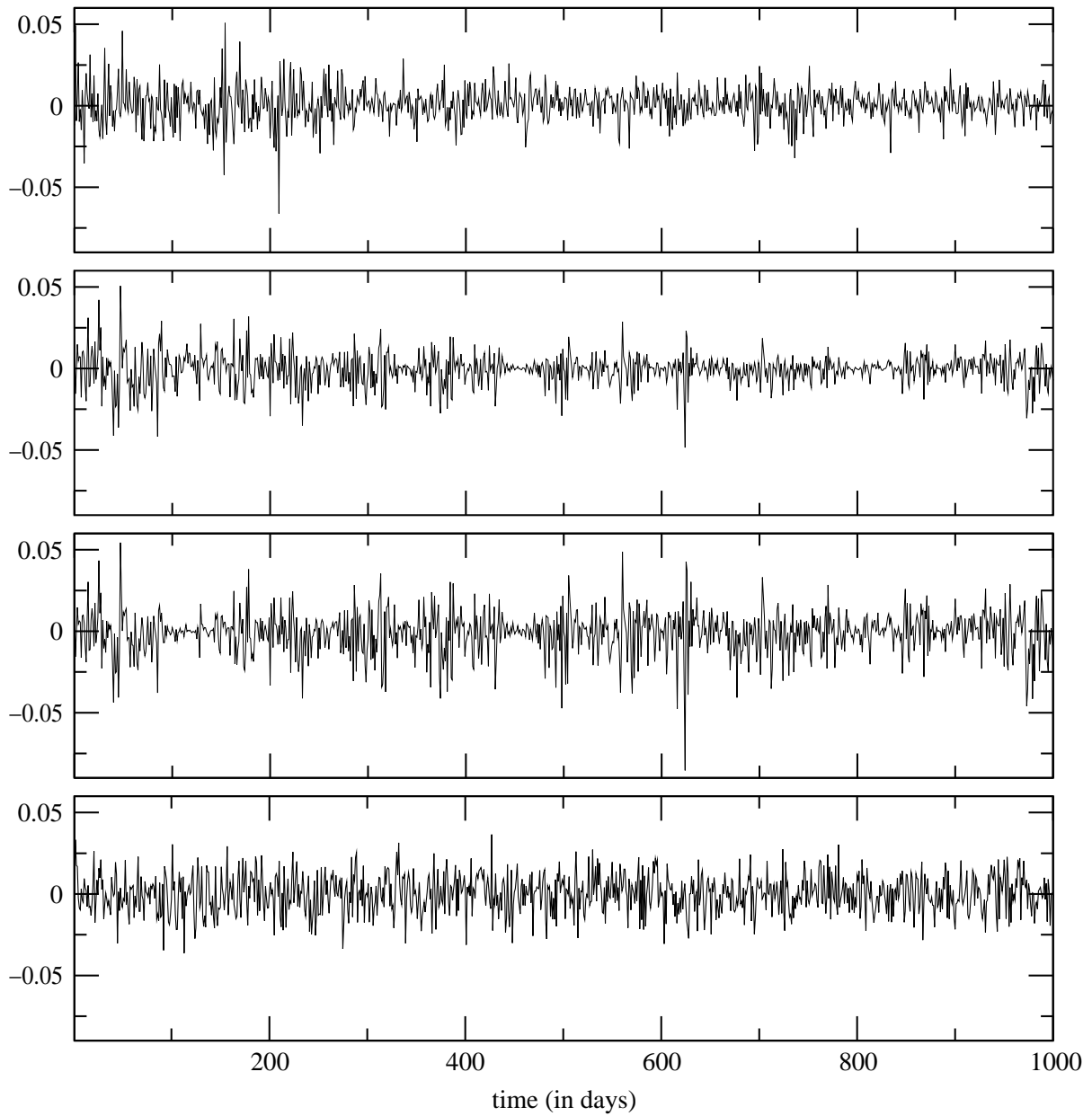}}
\caption{Dow-Jones index and simulation paths over 1000 days (approximately 4 years). Top 
figure shows the Dow-Jones index daily return changes over 1000 trading days and compare it 
with several simulations. From top to bottom we show random path of our three-dimensional 
stochstic volatility model~(\ref{3da})-(\ref{3d}), the path of the two-dimensional 
approach~(\ref{dx})-(\ref{sigsde}) and the Wiener process~(\ref{dx}) assuming constant $\sigma$. Parameters are given in Tables~\ref{momestim} and~\ref{corestim}.}
\label{comp-short}
\end{figure}

We have four more parameters to estimate, $\alpha$, $\alpha_0$, $\nu^2$ and $\rho$, 
and we will use both the leverage and the volatility correlations to obtain them. As 
we see in Figs.~\ref{corvol} and~\ref{djleverage}, the leverage correlation appears to be much more noisy than the volatility autocorrelation. For this reason we start with volatility autocorrelation for quantifying $\alpha$, $\alpha_0$, $\nu^2$. We 
subsequently analyze the leverage correlation for testing the estimated $\alpha$ and for finally obtaining the parameter  $\rho$.

As we have shown in the Section~\ref{4} there are different ways to interpret the 
volatility correlation from empirical data. In the previous Section our choice was to take 
$\Delta X(t)^2$ as the square of the daily volatility. We have computed the empirical 
volatility correlation and plotted it in Fig.~\ref{corvol} with the corresponding fit 
with our theoretical correlation in Eq.~(\ref{volcor3}):
\begin{eqnarray*}
\mbox{Corr}\left[dX(t)^2,dX(t+\tau)^2\right]=
N\biggl[a\bigl(2+ae^{-\alpha\tau}\bigr)e^{-\alpha\tau}+
b\bigl(2&+&be^{-\alpha_0\tau}\bigr)e^{-\alpha_0\tau}
+2abe^{-(\alpha+\alpha_0)\tau}\biggr],
\end{eqnarray*}
where $a=\nu^2-\lambda\hat{\nu}_0^2/(1-\lambda)$, $b=\hat{\nu}_0^2/(1-\lambda)$, 
$N=\left[1+8(\nu^2+\hat{\nu}_0^2)+4(\nu^2+\hat{\nu}_0^2)^2\right]^{-1}$. The fit 
plotted in Fig.~\ref{corvol} is done with the constraint that $a+b=(\nu^2+\hat{\nu}_0^2)=0.18$. 

The proposed fit is obtained by taking $1/\alpha_0\simeq 3$ years and $1/\alpha\simeq 10$ days. Recall that $\lambda=\alpha_0/\alpha$ is been assumed 
in previous Sections to be very small and this is now confirmed by empirical observations since $\lambda=1.3 \times 10^{-2}$. In such a case, we can consider that $a\simeq \nu^2$ and $b\simeq\nu_0^2$. The results obtained are given in 
Table~\ref{corestim} jointly with  $k$ and $k_0$ that are given by Eqs.~(\ref{nu0}) and (\ref{nu}).

The leverage effect provides a way to estimate the correlation coefficient $\rho$. 
It also serves us as a test of consistency in the estimation of the characteristic time $1/\alpha$. 
In a previous work~\cite{perello2}, 
it is been suggested that $1/\alpha\simeq 20$ days. However, 
the fit of the leverage function~(\ref{lev})--(\ref{levA}) to the Dow-Jones daily data 
(Fig.~\ref{djleverage}) is not very binding. The quality of this non linear fit remains basically unchanged for modifications in $1/\alpha$ inside the range $7-20$ days. For this reason we will keep our estimated $\alpha$ from volatility autocorrelation and accept that the results from the leverage correlation are consistent with the ones derived from the volatility correlation. 

Let us now focus on the estimation of $\rho$. Recall that our theoretical leverage is given
by Eqs.~(\ref{lev})-(\ref{levA}) with $a$ and $b$ given by Eq.~(\ref{ab}). 
When $\tau\rightarrow 0^+$, we have
\begin{equation}
{\cal L}(0^+)=\frac{2\rho\nu\sqrt{2\alpha}}{(1+\nu^2+\hat{\nu}_0^2)m},
\label{estimrho}
\end{equation}
which allows us to estimate $\rho$ once we know $\alpha=0.1 \mbox{ days}^{-1}$ and 
$(\nu^2+\hat{\nu}_0^2)=0.18$.  All these operations are summarized in Table~\ref{corestim}. 
The estimated magnitudes ${\cal L}(0^+)$ and $\alpha$ obtained from the Dow-Jones index are
of the same order as those given in~\cite{bouchaud} from an average over several stock indices. The value of $\alpha$ is also consistent with that found by 
Dragulescu and Yakovenko within the Heston model \cite{Yakov} (for which the volatility correlation function is a single exponential).

Finally, in Figs.~\ref{comp} and~\ref{comp-short} we simulate the resulting process 
with the parameters estimated above. Once we know the six estimated parameters, we 
can simulate the random dynamics for $\Delta X(t)$ and compare it with the empirical 
Dow-Jones time series. We also simulate the ordinary geometric Brownian motion where 
the volatility is constant together with the two-dimensional SV model studied in 
\cite{perello2} in which the normal level of volatility is constant. 
In Figs.~\ref{comp} and~\ref{comp-short} we see that the ordinary geometric
Brownian model cannot describe both the largest and the smallest fluctuations of 
daily returns. Furthermore, the two-dimensional SV model is not able to describe the 
clustering that appears when we take a longer number of days (see Fig.~\ref{comp}). 
Hence, we can conclude that SV models generate trajectories similar to that of the 
Dow-Jones but the three-dimensional model presented herein improves this description when 
we observe a wider time window. On the other hand, one should keep in mind that 
some aspects of the empirical data, such as the nearly log-normal distribution of 
volatility, or the extreme tails of the distribution of returns, 
are not reproduced by the present version of the model.

\section{Conclusions\label{7}}

In this paper, we have suggested a simple way to build a stochastic volatility 
model that accounts for the presence of 
multiple time scales in financial 
markets. The volatility correlation function contains at least a 
`short' time scale (tens of days) and a very long time scale that does not seem to
be present in the asymmetric return-volatility, which only exhibits the `short' time
scale. The idea is to let the `normal' level of volatility be itself time dependent
and evolve over long time scales.
The simplest setting is that of the linear Ornstein-Uhlenbeck model, where
the calculations are simple \cite{Stein,schobel,perello2}, and which can easily be calibrated 
to market data. 
The overall description of the volatility and the leverage correlation using this 
three dimensional model is
quite satisfactory, although several deficiencies remain: (a) the volatility 
process appears to need at least a third, very short time scale (on the order 
of a day), to be satisfactorily accounted for, (b) the empirical distribution
of volatility is poorly reproduced by the present Gaussian model, and (c) the assumption
that $dW_1$ is Gaussian does not allow one to account for extreme events. There are several 
natural extensions of the present model that could deal with these discrepancies:
(i) one could generalize the Heston model as is done here, such as to avoid 
the unrealistic `hump' of the volatility distribution close to zero; 
(ii) one could consider a model where the log-volatility, rather than the volatility, obeys 
an Ornstein-Uhlenbeck process, and (iii) one could build a `Matrioshka doll' 
Ornstein-Uhlenbeck process where the construction given here is iterated as many
times as needed to reproduce all the empirical time scales. In the limit of
an infinite number of time scales, it might be possible to construct in an
alternative way the multifractal random walk of Bacry, Muzy and Delour. 
Finally, it would be useful to extend the computation of the present paper 
to compute the conditional distribution of price changes over different time
scales, such as to obtain option prices. We hope to report the results of this 
computation in the near future.

\begin{acknowledgments}
This work has been supported in part by Direcci\'on General de Investigaci\'on under 
contract No. BFM2000-0795 and by Generalitat de Catalunya under contract No. 2000 SGR-00023. 
J-P B wants to thank M. Potters, A. Matacz, F. Selmi and V. Yakovenko for several useful
discussions.
\end{acknowledgments}

\appendix

\section{The Novikov theorem\label{apA}}

The Novikov theorem provides a way of evaluating averages involving the Wiener process and functionals of it. Suppose we have a well-behaved function of time 
$f(t;[\mbox{\boldmath$\xi$}])$ also depending functionally on an $n$-dimensional 
Gaussian white noise $\mbox{\boldmath$\xi$}(t)=(\xi_1(t),\cdots,\xi_n(t))$. 
The Novikov theorem (see Ref.~\cite{novikov} for further details) states that 
the nonlinear average  $\langle f(t;[\mbox{\boldmath$\xi$}]) \xi_j(t') \rangle$ 
can be written as
\begin{equation}
\langle f(t;[\mbox{\boldmath$\xi$}]) \xi_j(t') \rangle=\left\langle\frac{\delta
f(t;[\mbox{\boldmath$\xi$}])}{\delta \xi_j(t')}\right\rangle,\qquad(j=1,2,3,\cdots),
\label{novikov}
\end{equation}
where $\delta f(t;[\mbox{\boldmath$\xi$}])/\delta \xi_j(t')$ stands for the functional 
derivative of $f$ with respect to $\mbox{\boldmath$\xi$}$~\cite{hanggi}. We can write 
an alternative, and sometimes more convenient form of the theorem if we specify the 
correlation between the $n$ components of $\mbox{\boldmath$\xi$}(t)$:
$$
\langle \xi_i(t_1)\xi_j(t_2)\rangle =\rho_{ij}\delta(t_1-t_2).
$$
One can easily see that in this case we can write
\begin{equation}
\xi_i(t)=\xi(t)+\sum_{j\neq i}\rho_{ij}\xi_j(t),
\label{linear}
\end{equation}
where $\xi(t)$ is a Gaussian white noise independent of any other component of 
$\mbox{\boldmath$\xi$}(t)$, {\it i.e.}, $\langle \xi(t)\xi_j(t')\rangle =0$ for any 
$j\neq i$ and $t'$. On the other hand the chain rule applied to the functional derivative 
allows us to write~\cite{hanggi}
$$
\left\langle\frac{\delta f(t;[\mbox{\boldmath$\xi$}])}
{\delta\xi_j(t')}\right\rangle=\int dt'' \ \sum_{i=1}^n \left\langle
\frac{\delta f(t;[\mbox{\boldmath$\xi$}])}{\delta\xi_i(t'')} \ \frac{\partial\xi_i(t'')}
{\partial\xi_j(t')}\right\rangle.
$$
Then, taking into account Eq.~(\ref{linear}), the Novikov theorem can be set in the 
following alternative form:
\begin{equation}
\langle f(t;[\mbox{\boldmath$\xi$}])
\xi_j(t')\rangle=\sum_{i=1}^n\rho_{ij}\left\langle\frac{\delta
f(t;[\mbox{\boldmath$\xi$}])}{\delta \xi_i(t')}\right\rangle.
\label{novikov1}
\end{equation}
We are now able to prove  Eq.~(\ref{cornov}). In this  case we have to consider the average
$\langle \sigma(t)\sigma(t+\tau)^2\xi_1(t)\rangle$. Consequently
$f(t;[\mbox{\boldmath$\xi$}])=\sigma(t)\sigma(t+\tau)^2$ and the Novikov theorem, 
in the form given by
Eq.~(\ref{novikov}), tell us that
\begin{equation}
\langle \sigma(t)\sigma(t+\tau)^2
\xi_1(t)\rangle=\left\langle\frac{\delta
\left[\sigma(t)\sigma(t+\tau)^2\right]}{\delta \xi_1(t)}\right\rangle.
\label{novikov4}
\end{equation}
On the other hand we know from Eq.~(\ref{sigtot1}) that
\begin{equation}
\sigma(t)=m_0+k\int_{-\infty}^t
\xi_2(t')e^{-\alpha(t-t')}dt'+\frac{k_0}{1-\lambda}\int_{-\infty}^t
\xi_3(t')\left[e^{-\alpha_0(t-t')}-e^{-\alpha(t-t')}\right]dt',
\label{signov}
\end{equation}
where the integrals are represented in the It\^o sense. Observe that $\sigma(t)$ depends 
functionally on $\xi_2(t)$ and $\xi_3(t)$ but not on $\xi_1(t)$ and, as it is shown
by Eq.~(\ref{rho}), $\xi_1(t)$ and $\xi_2(t)$ are the only two correlated process 
with correlation coefficient $\rho$. Thus, the functional derivative reads
$$
\frac{\delta \left[\sigma(t)\sigma(t+\tau)^2\right]}{\delta
\xi_1(t)}=\rho\frac{\delta
\left[\sigma(t)\sigma(t+\tau)^2\right]}{\delta
\xi_2(t)},
$$
where we have used Eq.~(\ref{linear}) to see that $\partial\xi_2(t)/\partial\xi(t)_1=\rho$.
Therefore,
\begin{equation}
\frac{\delta\left[\sigma(t)\sigma(t+\tau)^2\right]}{\delta\xi_1(t)}=
\rho\left[\sigma(t+\tau)^2\frac{\delta\sigma(t)}{\delta\xi_2(t)}+2\sigma(t)\sigma(t+\tau)
\frac{\delta\sigma(t+\tau)}{\delta\xi_2(t)}\right].
\label{novikov5}
\end{equation}
The functional derivatives of $\sigma(t)$ and $\sigma(t+\tau)$ with respect to $\xi_2(t)$ 
can be obtained through Eq.~(\ref{signov}) after taking into account that $\xi_3(t)$ is 
independent of $\xi_2(t)$ and that $\delta\xi_2(t')/\delta\xi_2(t)=\delta(t-t')$~\cite{hanggi}.
Thus
\begin{equation}
\frac{\delta\sigma(t)}{\delta\xi_2(t)}=k\mbox{H}(0)=0,
\label{itonovikov1}
\end{equation}
since $H(0)=0$ ({\it cf.} Eq.~(\ref{heaviside})). Note that this is consistent with the 
It\^o interpretation which assumes that $\xi_2(t)$ lies outside of the integration 
interval $(-\infty,t)$ of the first integral on the rhs of Eq.~(\ref{signov}). Analogously
\begin{equation}
\frac{\delta\sigma(t+\tau)}{\delta\xi_2(t)}=ke^{-\alpha\tau}\mbox{H}(\tau).
\label{itonovikov2}
\end{equation}
Finally, from Eq~(\ref{novikov5}), we have
$$
\langle \sigma(t)\sigma(t+\tau)^2 \xi_1(t)\rangle=2\rho
ke^{-\alpha\tau}\langle\sigma(t)\sigma(t+\tau)\rangle\mbox{H}(\tau),
$$
which proves Eq.~(\ref{cornov}).

\section{Derivation of Equation~(\ref{volcor2}) \label{apB}}

We will demonstrate the equality
\begin{equation}
\frac{\langle dX(t)^2dX(t+\tau)^2\rangle-\langle dX(t)^2\rangle
\langle dX(t+\tau)^2\rangle}{\sqrt{\mbox{Var}[dX(t)^2]}
\sqrt{\mbox{Var}[dX(t+\tau)^2]}}=\frac{\langle\sigma(t)\sigma(t+\tau)\rangle^2
-\langle\sigma(t)\rangle^4}{4\langle\sigma(t)^2\rangle^2-3\langle\sigma(t)\rangle^4},
\label{b1}
\end{equation}
which is equivalent to prove Eq.~(\ref{volcor2}) from Eq.~(\ref{volcor}). To this end 
we will deal with each term on the left hand side of Eq.~(\ref{b1}) separately.

We first derive $\langle dX(t)^2dX(t+\tau)^2\rangle$. From Eq.~(\ref{dx}) we have
\begin{equation}
\langle dX(t)^2 dX(t+\tau)^2\rangle=
\langle\sigma(t)^2\sigma(t+\tau)^2dW_1(t)\xi_1(t)\rangle dt^2,
\label{b2}
\end{equation}
where we have assumed that $dW_1(t+\tau)$ is independent of the rest of stochastic variables 
and also that $\langle dW_1(t+\tau)^2\rangle=dt$. Since $dW_1(t)=\xi_1(t)dt$ we can write the
average on the rhs of Eq.~(\ref{b2}) as $\langle\sigma(t)^2\sigma(t+\tau)^2 
dW_1(t)\xi_1(t)\rangle$. Now the Novikov theorem,
Eqs.~(\ref{novikov}) and~(\ref{novikov1}), yields
\begin{eqnarray*}
\langle\sigma(t)^2\sigma(t+\tau)^2 dW_1(t)\xi_1(t)
\rangle=2\rho\left[\left\langle\frac{\delta\sigma(t)}
{\delta \xi_2(t)}\sigma(t)\sigma(t+\tau)^2dW_1(t)\right\rangle
\right.\left.+\left\langle\sigma(t)^2\sigma(t+\tau)
\frac{\delta\sigma(t+\tau)}{\delta \xi_2(t)} dW_1(t)\right\rangle\right]
\\
+\left\langle\sigma(t)^2\sigma(t+\tau)^2 \frac{\delta [dW_1(t)]}{\delta\xi_1(t)}\right\rangle,
\end{eqnarray*}
which, after using Eqs.~(\ref{itonovikov1})-(\ref{itonovikov2}), reads
\begin{eqnarray*}
\langle\sigma(t)^2\sigma(t+\tau)^2 dW_1(t)\xi_1(t)\rangle=
2\rho ke^{-\alpha\tau}\mbox{H}(\tau)
\bigl\langle\sigma(t)^2\sigma(t&+&\tau)dW_1(t)\bigr\rangle
+\left\langle\sigma(t)^2\sigma(t+\tau)^2
\frac{\delta [dW_1(t)]}{\delta\xi_1(t)}\right\rangle.
\end{eqnarray*}
In order to obtain an expression for the functional derivative
$\delta [dW_1(t)]/\delta\xi_1(t)$, we write $dW_1(t)$ in a somewhat intricate form
$$
dW_1(t)=\int_t^{t+dt}\xi_1(t')dt'.
$$
Thus
$$
\frac{\delta [dW_1(t)]}{\delta\xi_1(t)}=\int_t^{t+dt}\delta(t-t')dt'=1.
$$
Hence
\begin{equation}
\langle\sigma(t)^2\sigma(t+\tau)^2 
dW_1(t)\xi_1(t)\rangle=2\rho ke^{-\alpha\tau}\mbox{H}(\tau)\langle\sigma(t)\sigma(t+\tau)^2 
dW_1(t)\rangle+\langle\sigma(t)^2\sigma(t+\tau)^2\rangle.
\label{averagenovikov1}
\end{equation}
We apply again the Novikov theorem to the average $\langle\sigma(t)\sigma(t+\tau)^2 
dW_1(t)\rangle$ with the result
$$
\langle\sigma(t)\sigma(t+\tau)^2 dW_1(t)\rangle=
\rho\left\langle\frac{\delta}{\delta\xi_2(t)}[\sigma(t)^2\sigma(t+\tau)]\right\rangle dt,
$$
which, after using Eqs.~(\ref{itonovikov1})-(\ref{itonovikov2}), yields
$$
\langle\sigma(t)^2\sigma(t+\tau)dW_1(t)\rangle=
\rho ke^{-\alpha\tau}\mbox{H}(\tau)\langle\sigma(t)^2\rangle dt.
$$
Substituting this into Eq.~(\ref{averagenovikov1}) we have
$$
\langle\sigma(t)^2\sigma(t+\tau)^2 dW_1(t)\xi_1(t)\rangle=
2\rho^2k^2e^{-2\alpha\tau}\mbox{H}(\tau)\langle\sigma(t)^2\rangle dt
+\langle\sigma(t)^2\sigma(t+\tau)^2\rangle.
$$
Notice that due to the differential $dt$, the first term on the rhs of this 
equation is negligible in front of the second term. Thus
\begin{equation}
\langle\sigma(t)^2\sigma(t+\tau)^2 dW_1(t)\xi_1(t)\rangle=
\langle\sigma(t)^2\sigma(t+\tau)^2\rangle+\mbox{O}(dt),
\label{b3}
\end{equation}
and the substitution of Eq.~(\ref{b3}) into Eq.~(\ref{b2}) yields
\begin{equation}
\langle dX(t)^2 dX(t+\tau)^2\rangle=\left[\langle\sigma(t)^2\sigma(t+\tau)^2\rangle+
\mbox{O}(dt)\right]dt^2.
\label{vc1}
\end{equation}

On the other hand, since $dW_1(t)$ is independent of $\sigma(t)$ we can write
$$
\langle dX(t)^2\rangle\langle dX(t+\tau)^2\rangle=\langle \sigma(t)^2\rangle 
\langle \sigma(t+\tau)^2\rangle dt^2
$$
and, taking into account that process is in the stationary regime, we have
\begin{equation}
\langle dX(t)^2\rangle\langle dX(t+\tau)^2\rangle=\langle \sigma(t)^2\rangle^2 dt^2.
\label{vc2}
\end{equation}

Finally, we derive the variance:
$$
\mbox{Var}[dX(t)^2]=\langle \sigma(t)^4\rangle\langle dW(t)^4\rangle-\langle 
\sigma(t)^2\rangle^2\langle dW(t)^2\rangle^2,
$$
where we take into account that $\sigma(t)$ and $dW_1(t)$ are independent. 
We can go further since $\langle dW(t)^4\rangle=3\langle dW(t)^2\rangle^2=3dt^2$. Hence,
\begin{equation}
\mbox{Var}[dX(t)^2]=3\langle\sigma(t)^4\rangle dt^2-\langle\sigma(t)^2\rangle^2dt^2,
\label{vc3}
\end{equation}
and, due to the fact we are dealing with process in the stationary regime, 
the same result applies to the variance $\mbox{Var}[dX(t+\tau)^2]$.

Therefore, from Eqs.~(\ref{vc1}),~(\ref{vc2}) and~(\ref{vc3}), we have that 
correlation function given by Eq.~(\ref{b1}) is
\begin{equation}
\mbox{Corr}\left[dX(t)^2 dX(t+\tau)^2\right]=\frac{\langle\sigma(t)^2\sigma(t+\tau)^2\rangle-
\langle \sigma(t)^2\rangle^2}{3\langle\sigma(t)^4\rangle-\langle\sigma(t)^2\rangle^2}+
\mbox{O}(dt).
\label{volcor1}
\end{equation}
This expression contains two terms which must be also derived: 
$\langle\sigma(t)^2\sigma(t+\tau)^2\rangle$ and $\langle \sigma(t)^4\rangle$. 
Again, these terms can be obtained with the help of Novikov theorem. 
Lengthy and tedious calculations allow us write them in terms of other 
expressions presented above ({\it cf.} Eqs.~(\ref{varsig}) and~(\ref{corsigma})). 
The first one is
$$
\langle\sigma(t)^2\sigma(t+\tau)^2\rangle=2\left[\langle\sigma(t)\sigma(t+\tau)\rangle^2-
\langle\sigma(t)\rangle^4\right]+\langle\sigma(t)^2\rangle^2,
$$
while the second one is obtained from the first one by assuming $\tau=0$. That is:
\begin{equation}
\langle\sigma(t)^4\rangle=3\langle\sigma(t)^2\rangle^2-2\langle\sigma(t)\rangle^4,
\label{varsig2}
\end{equation}
If we insert these averages to the correlation~(\ref{volcor1}), we prove the equality
given by Eq.~(\ref{b1}) and finally obtain Eq.~(\ref{volcor2}).

\section{An alternative noise correlation \label{apC}}

\begin{figure}[t,b]
\centerline{\epsfig{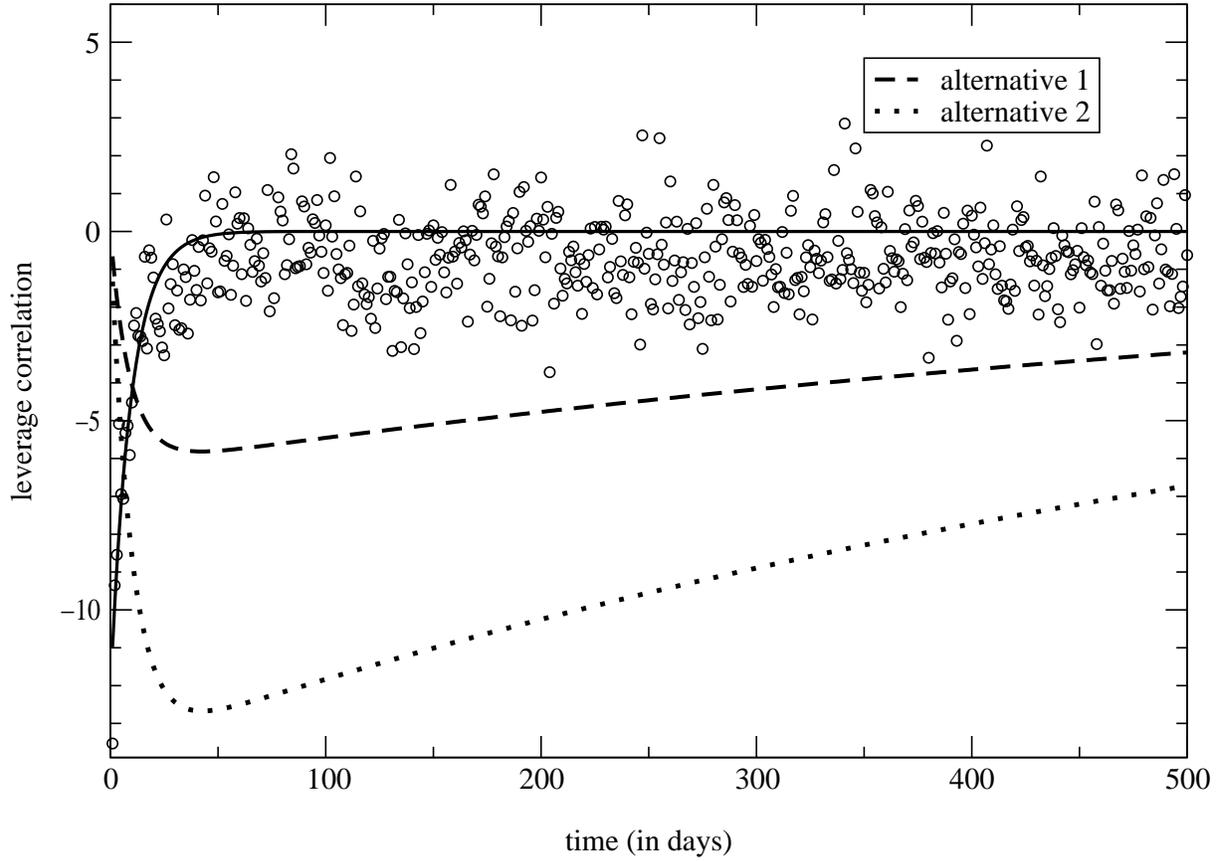}}
\caption{The leverage effect in the Dow-Jones daily index with the alternative leverage. 
We plot the leverage function ${\cal L}(\tau)$ and $\bar{\cal L}(\tau)$ and compare with 
the Dow-Jones index from 1900 until 2000. The fit with the leverage function ${\cal L}(\tau)$
given by Eq.~(\ref{lev}) takes parameters given by Table~\ref{corestim}. We also plot 
the two alternative leverage functions given by Eq.~(\ref{lev-c}). Alternative 1 shows 
the possible leverage with $\alpha=0.1$, $a=0.14$, $\alpha_0=0.04$ and $b=0.04$ while 
alternative 2 takes parameters $\alpha_0=0.1$, $b=0.14$, $\alpha=0.04$ and $a=0.04$.}
\label{alterleverage}
\end{figure}
We recall that our model is given by a three-dimensional random process~(\ref{3da})--(\ref{3d})
\begin{eqnarray*}
&&dX(t)=\sigma(t)dW_1(t), \\
&&d\sigma(t)=-\alpha[\sigma(t)-m(t)]dt+kdW_2(t),\\
&&dm(t)=-\alpha_0[m(t)-m_0]dt+k_0dW_3(t),
\end{eqnarray*}
where $dW_i(t)=\xi_i(t)dt$ $(i=1,2,3)$ are Wiener processes, {\it i.e.,} $\xi_i(t)$ are 
zero-mean Gaussian white noises with cross-correlation given by
$
\langle\xi_i(t)\xi_j(t')\rangle=\rho_{ij}\delta(t-t').
$
Note that $\rho_{ij}=\rho_{ji}$ and $\rho_{ii}=1$. We have supposed that $W_3(t)$ is 
independent of $W_1(t)$ and $W_2(t)$. And we also have assumed that there exists a non 
zero correlation between $W_1(t)$ and $W_2(t)$ that was quantified by the parameter $\rho$.

This appendix modifies this framework by changing the existing correlations betweeen the 
Wiener processes $W_i$ ($i=1,2,3$). We want to analyze the consequences on the model 
when we allow correlations between $W_1(t)$ and $W_3(t)$ instead of between $W_1(t)$ and 
$W_2(t)$. In this case, the correlation matrix is
\begin{equation}
(\rho_{ij})=
\left(\begin{array}{ccc}
1 & 0 & \rho \\
0 & 1 & 0 \\
\rho & 0 & 1
\end{array}\right)
\label{rho-app}
\end{equation}
where, again, $-1\leq\rho\leq 1$.

One can easily see that this change in the correlation matrix does not affect most 
of the main statistical properties of the model although there is one important property 
that is  modified: the leverage correlation.

According to the definition~(\ref{leverage}), the leverage correlation is quantified by
\begin{equation}
{\cal L}(\tau)\equiv
\frac{1}{Z}\langle dX(t+\tau)^2dX(t)\rangle
\label{leverage-app}
\end{equation}
where $X(t)$ is the zero-mean return and 
$Z=\langle dX(t)^2\rangle^2=m_0^4(1+\nu^2+\widehat{\nu}_{0}^2)^2dt^2$ 
({\it cf.} Eq.~(\ref{z})). We need to calculate the correlation
$$
\langle dX(t+\tau)^2dX(t)\rangle=
H(\tau)\langle\sigma(t)dW_1(t)\sigma(t+\tau)^2dW_1(t+\tau)^2\rangle.
$$
Its derivation is similar to the one given in Section~\ref{3}. 
Since we follow the It\^o convention, $dW_1(t+\tau)$ is independent of the 
rest of variables when $\tau> 0$. This fact and the averages $\langle dW_1(t)\rangle=0$ 
and $\langle dW_1(t)^2\rangle=dt$ allow us to show that
$$
\langle dX(t+\tau)^2dX(t)\rangle=\langle\sigma(t)\sigma(t+\tau)^2dW_1(t)\rangle H(\tau) dt,
$$
where $H(\tau)$ is the Heaviside step function given by Eq.~(\ref{heaviside}). 
The next step of the calculation needs to apply the Novikov theorem enunciated in 
Appendix~\ref{apA}, although now its result appears to be different from the one 
given by Eq.~(\ref{cornov}) since the correlation is between $W_1(t)$ and $W_3(t)$. 
In the present case the Novikov theorem given by Eq.~(\ref{novikov1}) allows us to write
\begin{equation}
\langle\sigma(t)\sigma(t+\tau)^2dW_1(t)\rangle=
\sum_{i=1}^3\rho_{1i}
\left\langle\frac{\delta[\sigma(t)\sigma(t+\tau)^2]}{\delta \xi_i(t')}\right\rangle dt.
\label{novikov1-c}
\end{equation}
On the other hand we know from Eq.~(\ref{sigtot1}) that
\begin{equation}
\sigma(t)=m_0+k\int_{-\infty}^t
\xi_2(t')e^{-\alpha(t-t')}dt'+\frac{k_0}{1-\lambda}\int_{-\infty}^t
\xi_3(t')\left[e^{-\alpha_0(t-t')}-e^{-\alpha(t-t')}\right]dt',
\label{signov-c}
\end{equation}
where the integrals are represented in the It\^o sense. Since volatility~(\ref{signov-c}) 
does not depend on $\xi_1$ and the only two non zero contributions correspond to $\rho_{11}=1$ 
and $\rho_{13}=\rho$ ({\it c.f.} Eq.~(\ref{rho-app})),
\begin{equation}
\langle\sigma(t)\sigma(t+\tau)^2dW_1(t)\rangle=2\rho\left\langle\sigma(t)\frac{\delta
\sigma(t+\tau)}{\delta \xi_3(t')}\right\rangle dt.
\label{novikov2-c}
\end{equation}
But from Eq.~\ref{signov-c}
\begin{equation}
\frac{\delta\sigma(t+\tau)}{\delta\xi_3(t)}=\frac{k_0}{1-\lambda}\left(e^{-\alpha_0\tau}-
e^{-\alpha\tau}\right)\mbox{H}(\tau),
\label{itonovikov2-c}
\end{equation}
and $\langle\sigma(t)\rangle=m_0$. Hence the leverage correlation given by 
Eq.~(\ref{leverage-app}) reads
\begin{equation}
\bar{\cal L}(\tau)=\mbox{H}(\tau)\bar{A}(\tau)(e^{-\alpha_0\tau}-e^{-\alpha\tau}),
\label{lev-c}
\end{equation}
where
\begin{equation}
\bar{A}(\tau)\equiv \frac{2\rho\nu_0\sqrt{2\alpha_0}}{m_0(1-\lambda)
\left(1+\nu^2+\hat{\nu}_0^2\right)^2}\left[1+\left(\nu^2-\frac{\hat{\nu}_0^2\lambda}{1-\lambda}\right)e^{-\alpha\tau}+\frac{\hat{\nu}_0^2}{1-\lambda}e^{-\alpha_0\tau}\right].
\label{levAc}
\end{equation}

We now compare the two leverage correlations ${\cal L}(\tau)$ and 
$\bar{\cal L}(\tau)$. In Section~\ref{3}, we obtained ${\cal L}(\tau)$ 
given by Eqs.~(\ref{lev})--(\ref{levA}) and see that the limit ${\cal L}(0^+)$ 
goes to a certain negative value. In contrast, in the present alternative model, 
$\bar{\cal L}(0^+)=0$. In Fig.~\ref{alterleverage}, we see that leverage of the 
Dow-Jones index goes to $-11.9$ when $\tau\rightarrow 0^+$. Hence, this alternative model 
is in clear contradiction with empirical observations. In addition, we also observe in 
Fig.~\ref{alterleverage} that $\bar{\cal L}(\tau)$ has a longer range than the one observed
in the Dow-Jones index. The reason for this being is that the exponential time decay of the leverage function~(\ref{levAc}) is dominated by the same parameter as to the case of the volatility autocorrelation~(\ref{volcor3}). Thus, in the present case, leverage and volatility correlations have the same long range persistence and this is also in clear contradiction with the empirical results telling us that leverage has a much shorter memory than the volatility (see and compare Figs.~\ref{corvol} and~\ref{alterleverage}).

\end{document}